\documentclass{article}
\usepackage{spconf,amsmath,graphicx}
\usepackage{url}
\usepackage[caption=false,font=footnotesize]{subfig}
\usepackage[dvipsnames]{xcolor}
\usepackage{hyperref}
\usepackage{colortbl}

\title{Transferring neural speech waveform synthesizers to musical instrument sounds generation}

\name{Yi Zhao$^{\star}$ \qquad Xin Wang$^{\star}$ \qquad Lauri Juvela$^{\dagger}$ \qquad Junichi Yamagishi$^{\star}$}

            \address{$^{\star}$ National Institute of Informatics, Tokyo, Japan \\
                $^{\dagger}$ Department of Signal Processing and Acoustics, Aalto University, Finland}

\begin{document}
\ninept
\maketitle
\begin{abstract}
Recent neural waveform synthesizers such as WaveNet, WaveGlow, and the neural-source-filter (NSF) model have shown good performance in speech synthesis despite their different methods of waveform generation. The similarity between speech and music audio synthesis techniques suggests interesting avenues to explore in terms of the best way to apply speech synthesizers in the music domain.
This work compares three neural synthesizers used for musical instrument sounds generation under three scenarios: training from scratch on music data, zero-shot learning from  the speech domain, and fine-tuning-based adaptation from the speech to the music domain. 
The results of a large-scale perceptual test demonstrated that the performance of three synthesizers improved when they were pre-trained on speech data and fine-tuned on music data, which indicates the usefulness of knowledge from speech data for music audio generation. Among the synthesizers, WaveGlow showed the best potential in zero-shot learning while 
NSF performed best in the other scenarios and could generate samples that were perceptually close to natural audio.

\end{abstract}
\section{Introduction}
\label{sec:intro}

Many technological parallels can be drawn between the synthesis of speech and musical instruments, both historically and in the present deep learning era.
Previously, concatenative techniques had been widely applied in text-to-speech (TTS)~\cite{Hunt1997-unit-selection-tts} and musical sound synthesis~\cite{schwarz2006concatenative}.
Although such approaches can produce high-quality waveforms by re-arranging and concatenating small pre-recorded waveform segments,
they are inflexible and require a large footprint to store the sample library.
The speech and musical sound synthesis fields also have a strong tradition in parametric synthesis models, 
which enable waveform generation from a compact set of acoustic control features. 
In speech, parametric models have been combined with data-driven techniques to create statistical parametric speech synthesis (SPSS) systems~\cite{zen2009-spss}.  
In musical sound, although parametric synthesizers have a wide range of artistic uses \cite{bode1984history}, data-driven reproduction of instrument audio by parametric models is still difficult \cite{beauchamp2007analysis}. 

More recently, deep learning techniques have reshaped the way speech synthesis is done. 
Many neural waveform synthesizers surpass traditional parametric synthesis models in speech quality~\cite{tamamori2017speaker,prenger2019waveglow}. 
These waveform synthesizers avoid many speech-specific assumptions by using generic neural networks, e.g., the convolution net in WaveNet ~\cite{tamamori2017speaker} and recurrent net in SampleRNN~\cite{mehri2016samplernn}. 
%

Further, musical sound synthesis also benefits from neural generative models. 
Related works have focused on either unconditional generation of continuous music~\cite{dieleman2018-modelling-raw-audio-at-scale} or conditional synthesis of isolated musical notes~\cite{engel2017neural, engel2019-gansynth}. 
Meanwhile, approaches resembling the speech synthesis have been proposed for multi-instrument music generation~\cite{kim2019neural}, where a WaveNet generates audio from the Mel-spectrograms. A similar approach 
has also been proposed for singing synthesis~\cite{Blaauw2019-voice-cloning-neural-singing-synthesis}. 

The similarities between musical and speech modeling technologies suggest the possibility of a universal neural waveform synthesizer that is capable of synthesizing both speech (independent of speaker) and musical audio (independent of instrument), as well as the possibility for musical audio synthesis to benefit from speech models.
In this paper, we investigate transferring neural speech waveform synthesizers to musical instrument sounds generation under three scenarios: 1) generating the sounds with synthesizers trained only on music data; 2) generating the sounds from synthesizers trained on speech data  directly, i.e., zero-shot learning; and 3) generating the sounds after fine-tuning the pre-trained speech synthesizer on music data.
We selected WaveNet and WaveGlow~\cite{prenger2019waveglow} for experiments due to their superior performance
to speech synthesis~\cite{govalkarcomparison}.
We also included the recently proposed NSF  synthesizer~\cite{wang2018neural,wang2019neural}, which works in a different way from WaveNet and WaveGlow.
Results of a large-scale perceptual test suggest that fine-tuning speech neural synthesizers on music data can improve the quality of the synthesized audio. Among the three types of synthesizers, WaveGlow showed better potential in the zero-shot learning scenario. NSF surpassed the other models in the fine-tuning and training-from-scratch cases, thanks to the source-filter model assumption.

Section 2 explains the details of the selected neural waveform synthesizers and reconstruction methods. Section 3 discusses the experiments and results. We close in section with a brief summary and mention of future work.

\section{Neural waveform synthesizers and training methods}
\label{sec:neu_re}

\subsection{Selected neural waveform synthesizers} 
\label{ssec:neu}
We select WaveNet, WaveGlow, and NSF and compare their performance on reconstructing the sounds of musical instruments. We choose these three because each represents a different neural waveform modeling approach: WaveNet for the auto-regressive (AR) neural approach, WaveGlow for the inverse-AR-flow-based approach, and NSF for the third approach, which uses neither AR nor inverse-AR.

WaveNet~\cite{oord2016wavenet} is the earliest neural waveform synthesizer for raw audio waveform generation. It is a dilated convolutional neural network that generates each waveform sample based on previous ones. The stack of dilated causal convolution layers gives WaveNet a large receptive field that grows exponentially as the number of convolution layers increases. Although WaveNet produces high-quality speech waveforms, its sequential AR generation process is highly time-consuming.

WaveGlow~\cite{prenger2019waveglow} was inspired by a neural image generator called Glow~\cite{kingma2018glow}. On the basis of the inverse-AR flow framework ~\cite{NIPS2016_6581}, WaveGlow transforms the target waveform into a noise sequence and maximizes its likelihood over a Gaussian distribution during training. For generation, WaveGlow draws a noise sequence and de-transforms it back into the waveform domain. WaveGlow has an advantage over WaveNet in that it can produce waveforms in real time. However, the training time is prohibitively long due to the sequential transformation defined in the inverse-AR flow framework.

The Neural source-filter (NSF) model uses
dilated-convolution-based filter modules to transform sine-based excitation into an output waveform, where the frequency of the excitation is determined by the input F0~\cite{wang2018neural}. 
Unlike WaveNet and WaveGlow, NSF does not use AR or inverse-AR flow. It was demonstrated on a Japanese speech corpus that NSF performed similarly to WaveNet in terms of speech quality while being faster in generation. 
For music signal generation, one potential advantage of NSF is that the F0 of the generated waveform is expected to be consistent with the input F0. This is due to the source-filter structure of NSF, where the sine source excitation carries the input F0 and the neural filter transforms the excitation into the output waveform without tampering with its pitch structure.

\subsection{Three scenarios of model training }
\label{ssec:re}

We evaluate the performance of the three neural waveform synthesizers in three scenarios: training from scratch, zero-shot learning and fine-tuning. 
In the first scenario, we train the neural waveform synthesizers from scratch on a limited amount of music data and examine whether these generators can reconstruct musical signals.
In the second one, we define a zero-shot learning scenario where the neural waveform synthesizers trained on speech data are directly used to reconstruct music signals. The purpose here is to examine whether neural generators trained on speech data can be generalized to music signal generation.
In third scenario, we use music data to fine-tune the neural waveform synthesizers, which are pre-trained on speech data. 

In all of these scenarios, the waveform synthesizers produce music signals given input natural acoustic features such as Mel-spectrogram and F0. By comparing the performance of the synthesizers in the three scenarios, we hope to clarify not only the performance of each waveform synthesizer on music signal modeling but also how a model pre-trained on speech data can be used for music signal generation.

\begin{table}[!t]
\caption{Statistics of training, development, and test sets extracted from URMP corpus. \textit{Dur} denotes duration in minutes. $\text{F0}_\text{max}$ and $\text{F0}_\text{min}$ denote maximum and minimum F0 values in Hz, respectively. Note that music pieces in test set were different from those in training set.}
\begin{center}
\footnotesize
\setlength{\tabcolsep}{2pt}
\scalebox{0.93}{
\begin{tabular}{c|c|c|rr|c|c|rr|c|c|rr}
\hline\hline
      & \multicolumn{4}{c|}{Training set} & \multicolumn{4}{c|}{Development set} & \multicolumn{4}{c}{Test set} \\
      & \multicolumn{4}{c|}{110 music pieces} & \multicolumn{4}{c|}{22 music pieces} & \multicolumn{4}{c}{17 music pieces} \\
\cline{2-13}
     & \# & \textit{Dur}. & $\text{F0}_\text{max}$ & $\text{F0}_\text{min}$ 
     & \# & \textit{Dur}. & $\text{F0}_\text{max}$ & $\text{F0}_\text{min}$
     & \# & \textit{Dur}. & $\text{F0}_\text{max}$ & $\text{F0}_\text{min}$ \\
\hline
Violin & 26 & 47.8 & 1417 & 194 & 5 & 11.0 & 1436 & 194 & 3 & 2.1 & 707 & 191 \\
Cello & 10 & 23.0 & 694 & 111 & 2 & 2.2 & 556 & 160 & 1 & 0.8 & 336 & 173 \\
Viola & 6 & 12.9 & 295 & 62 & 4 & 7.8 & 302 & 61 & 1 & 1.7 & 355 & 62 \\
Bass & 3 & 6.1 & 225 & 55 & – & – & – & – & – & – & – & – \\
Flute & 13 & 26.0 & 1907 & 273 & 2 & 3.2 & 1791 & 291 & 3 & 3.7 & 1713 & 293 \\
Oboe & 5 & 10.0 & 949 & 236 & – & – & – & – & 1 & 1.7 & 1208 & 392 \\
Clarinet & 8 & 14.3 & 891 & 141 & 1 & 2.2 & 886 & 164 & 1 & 1.5 & 732 & 144 \\
Saxophone & 9 & 12.8 & 905 & 147 & 1 & 1.2 & 604 & 244 & 1 & 1.7 & 443 & 174 \\
Bassoon & 2 & 3.6 & 397 & 97 & 1 & 0.7 & 297 & 129 & – & – & – & – \\
Trumpet & 16 & 33.1 & 966 & 149 & 3 & 4.5 & 641 & 165 & 3 & 3.5 & 844 & 203 \\
Horn & 4 & 9.3 & 474 & 121 & – & – & – & – & 1 & 0.9 & 527 & 184 \\
Trombone & 5 & 12.1 & 421 & 82 & 2 & 4.2 & 370 & 65 & 1 & 0.9 & 267 & 84 \\
Tuba & 3 & 3.8 & 213 & 63 & 1 & 4.2 & 214 & 79 & 1 & 0.9 & 130 & 43 \\
\hline\hline
\end{tabular}
}
\end{center}
\label{tab:corpus}
\end{table}

\section{Experiments and results}
\label{sec:exp}
\subsection{Database and protocols}
\label{ssec:dbase}

We used the University of Rochester Music Performance (URMP) database~\cite{li2018creating} for experiments. Although the URMP database is designed for audio-visual analysis of musical performances, we selected it because it provides recordings of musical instruments in separate tracks and manually corrected F0 trajectories at the frame level (10 ms). The database covers 13 instruments including violin, clarinet, and trumpet.

We used 149 tracks of recordings from the database, for a total duration of around 4.5 hours. Each track has only one instrument recorded with a sampling rate of 48 kHz and 26 bits in mono-channel. Because the 149 tracks were performed as 44 music ensembles (11 duets, 12 trios, 14 quartets, and seven quintets), to avoid overlapping of the music scores, we used tracks from disjoint music pieces to train and test the waveform synthesizers. Specifically, 132 tracks from 38 music pieces were used for the training (110 tracks, 3.5 hours) and development sets (22 tracks, 0.6 hours). The 17 tracks from the remaining six music pieces, the duration of which was around 20 minutes, were used for testing. All the instruments that appear in the test set are covered by the training set, i.e., no unseen instruments appear in the test set\footnote{Because every instrument appeared in multiple music pieces, it impossible to find one music piece with an unseen instrument.}. The statistics for each set are listed in Table~\ref{tab:corpus}.

For zero-shot learning and fine-tuning, we selected 87 speakers' data from the VCTK speech corpus \cite{veaux2017cstr} to pre-train the neural waveform synthesizers. The number of speech utterances was 17,400, with the duration of around 21 hours.

\subsection{Feature configurations}
\label{ssec:features}
We extracted the Mel-spectrogram from the natural signals and used it as the input feature to train and test the neural waveform synthesizers. For training from scratch, we down-sampled the original signals into 24 kHz and extracted the Mel-spectrogram with a frame shift of 120 waveform sampling points (5 ms). The dimension of the Mel-spectrum in one frame was 80. We used this configuration because it is widely adopted for speech synthesis tasks~\cite{wang2018comparison}. For zero-shot learning and fine-tuning, we down-sampled the audio files into 22.05 kHz and extracted 80-dimensional Mel-spectra at a frame shift of 256 points (approx.\ 11.6 ms). This configuration is the same as that used in the pre-trained WaveGlow model\footnote{\url{https://github.com/NVIDIA/waveglow}\label{fn:WG}}. Therefore, we used this configuration for WaveGlow and the other two models in the case of zero-shot adaption and fine-tuning.

In addition to the Mel-spectrogram, we fed the ground-truth F0 sequence to WaveNet and NSF. The F0 sequence was up-sampled by replication so that its length was equal to the number of frames in the Mel-spectrogram. We did not use F0 in WaveGlow because the pre-trained WaveGlow model does not use it. Since the high-resolution mel spectrograms carries F0 information, we think it still fair to compare the three waveform synthesizers regardless of the F0 input.

\subsection{Model configurations}
\label{ssec:config}
\begin{figure}[!t]
 \centering
\includegraphics[width=0.97\columnwidth]{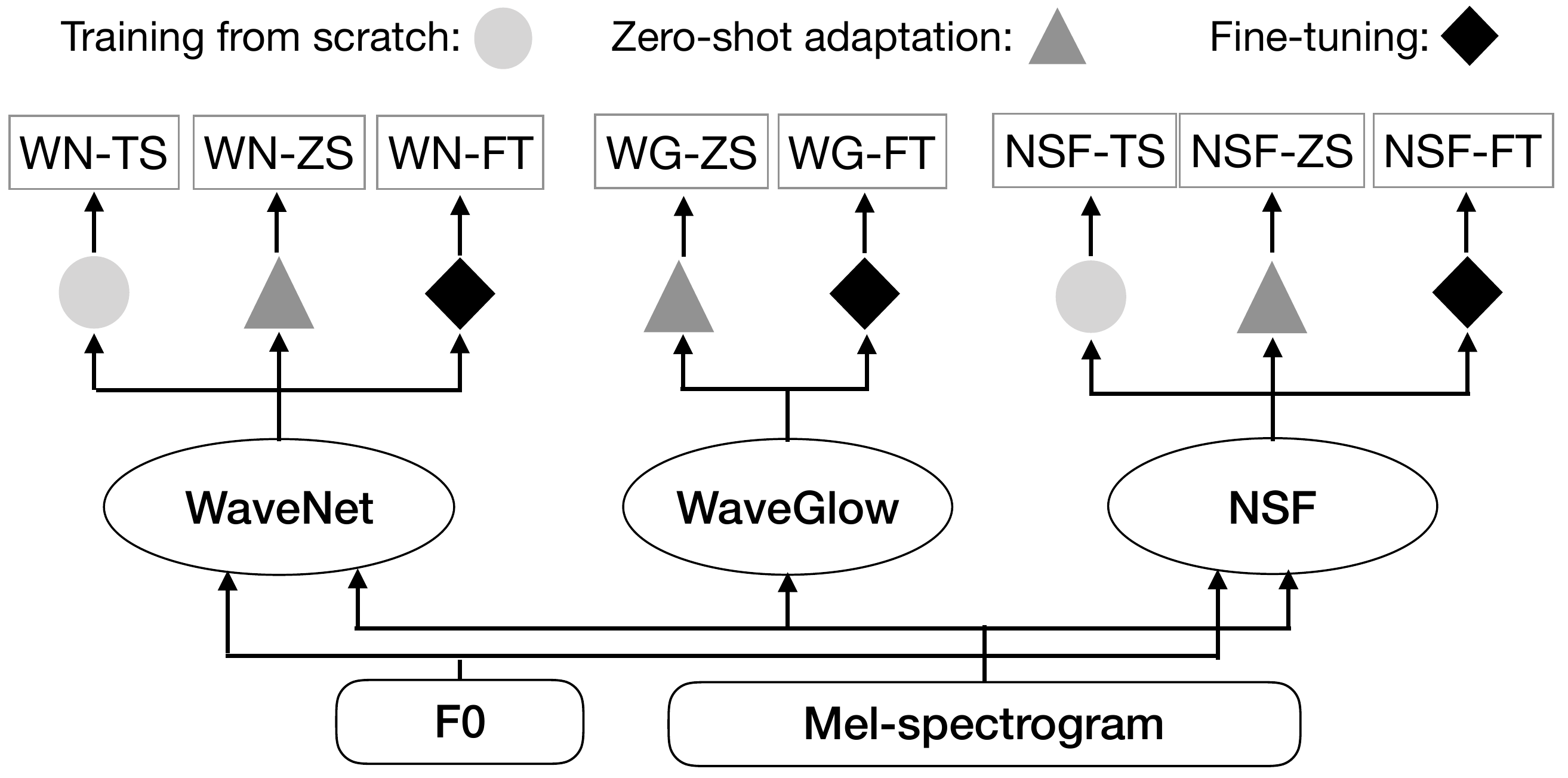}
\caption{Flowchart for the eight strategies compared in this study. }
\label{fig:sys}
\end{figure}

We compared WaveNet, NSF, and WaveGlow in the three scenarios defined in Section~\ref{ssec:re}. This testing framework is plotted in Figure~\ref{fig:sys}. Note that the WaveGlow trained from scratch was excluded from the experiment because it generated nothing but random noises, even though we had trained the model for more than one week.

\noindent\textbf{WaveNet}: The WaveNet was configured in a similar manner to that in our previous work on speech synthesis~\cite{wang2018comparison}. 
Its backbone consists of a linear projection input layer that takes the feedback waveform as input, 30 blocks of dilated convolution, and a post-processing block that computes the categorical distribution for 10-bit quantized and $\mu$-law companded waveform values. The $k$-th dilated convolution block has a dilation size of $2^{\mathrm{mod}(k-1,10)}$, where $\mathrm{mod}(\cdot)$ is modulo operation. Similar to the original design \cite{oord2016wavenet}, each dilated convolution block contains one dilated convolution layer, a gated activation function, and linear transformation layers. The conditional acoustic features are pre-processed by a bi-directional recurrent layer with 32 long short-term-memory units (LSTM) in both directions and a 1-D convolution layer with a kernel size of three and 63 output channels. The 1-dimensional F0 input is then concatenated with the output of the 1-D convolution layer, and the 64-dimensional conditional features are up-sampled by repeating the values and are then fed to each dilated convolution block.

\noindent\textbf{WaveGlow}: Our WaveGlow model followed the standard configuration in~\cite{prenger2019waveglow}. All related experiments are based on Nvidia's implementation described earlier.

\noindent\textbf{NSF}: The NSF model, or more accurately the harmonic-plus-noise NSF model with trainable maximum-voice-frequency (MVF), was also built using a recipe for speech synthesis \cite{wang2019neural}. It contains a source module, a neural filter module, and a conditional module. The conditional module is similar to that in the WaveNet except that one bi-directional LSTM and a 1-D convolution layer are added to predict the MVF \cite{wang2019neural}. The conditional module up-samples the F0 trajectory and feeds it to the source module. At the same time, it feeds the processed acoustic features to the neural filter module. The source module takes the F0 input and generates a sine waveform whose instantaneous frequency is equal to the F0 value. For time steps without F0, the source excitation is Gaussian noise. 
Given the excitation, the neural filter modules use five dilated convolution blocks to produce a waveform. At the same time, it uses a separate block to transform a noise sequence into another waveform. The two waveforms are then processed by a low-pass and a high-pass finite-impulse-response filter, respectively, and the sum of the two waveforms is the final output waveform. The cut-off frequency of the two FIR filters is determined by the predicted MVF. Note that NSF generates continuous-valued waveforms rather than quantized ones.

Both WaveNet and the NSF model are implemented on a modified CURRENNT toolkit \cite{weninger2015introducing}. The training recipes can be found online.

\begin{figure*}[!t]
\centering
\subfloat{\includegraphics[width=1.0\textwidth]{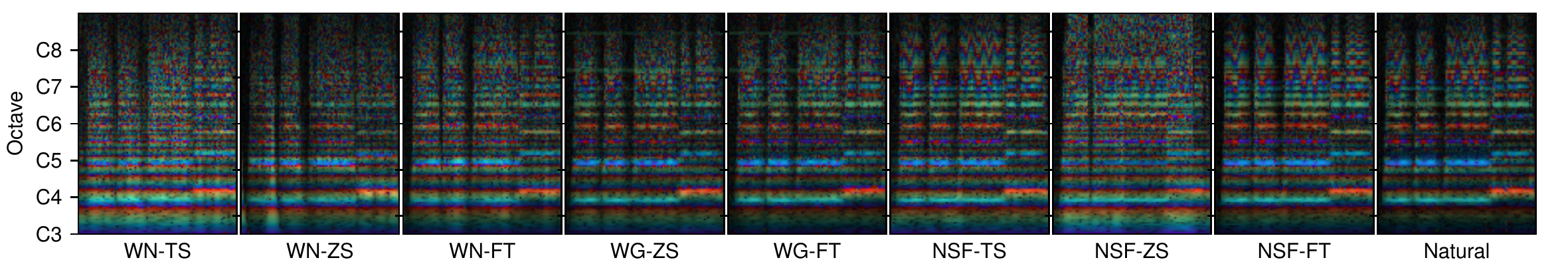}
}
\vspace{-3mm}
\subfloat{\includegraphics[width=1.0\textwidth]{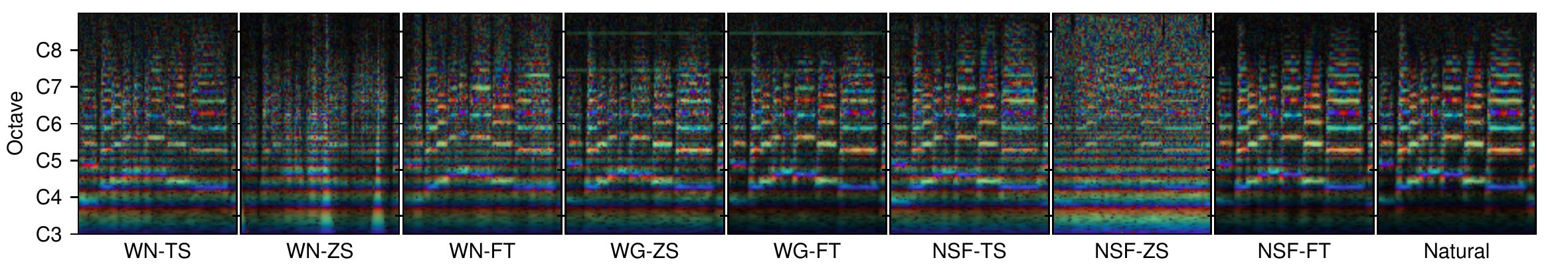}
}
\vspace{-3mm}
\caption{Rainbow-gram of natural and generated music signals for violin (1st row) and trumpet (2nd row).} 
\label{fig:spec_rainbow}
\vspace{-2mm}
\end{figure*}

\subsection{Conditions for evaluation}
\label{ssec:eva}
We followed Figure~\ref{fig:sys} and got eight experimental models: WaveNet-scratch (\texttt{WN-TS}), WaveNet-zero-shot (\texttt{WN-ZS}), WaveNet-fine-tune (\texttt{WN-FT}), NSF-scratch (\texttt{NSF-TS}), NSF-zero-shot (\texttt{NSF-ZS}), NSF-fine-tune (\texttt{NSF-FT}), WaveGlow-zero-shot (\texttt{WG-ZS}), and WaveGlow-fine-tune (\texttt{WG-FT}). 
We also used natural sounds (\texttt{NAT}) as reference. Thus we had nine experimental groups to evaluate.

A perceptual evaluation was organized on a crowd-sourcing platform to evaluate the quality of the generated waveforms\footnote{Audio samples are available at: \url{https://nii-yamagishilab.github.io/samples-nsf/neural-music.html}}
and their similarity to the natural waveforms in terms of instrument timbre. 
To reduce the burden on the crowdworkers, the generated waveforms in the test set were manually split into segments of at most 15 seconds in duration. Each experimental model consisted of 150 music signal segments, and 1350 segments were used for perceptual evaluation ($1350 = 150  \times 9$ systems). Each sample was evaluated six times in order to avoid human bias. The actual number of listeners who participated in our test was 136. Note that the listeners were not required to be music professionals.

To evaluate quality, listeners were asked to rate how natural each sample sounded on a five-point scale, where 1 denotes being completely unnatural and 5 completely natural. For similarity, listeners were asked to ignore the content of the music and concentrate only on the instrument characteristics. Samples and the corresponding natural sample were presented in pairs at every turn and listeners were asked to rate on a five-point scale, where 1 means absolutely different and 5 means absolutely the same.

\subsection{Results and discussion}
\label{sec:res}

\begin{figure}[thb]
 \centering
\includegraphics[width = 8.6cm]{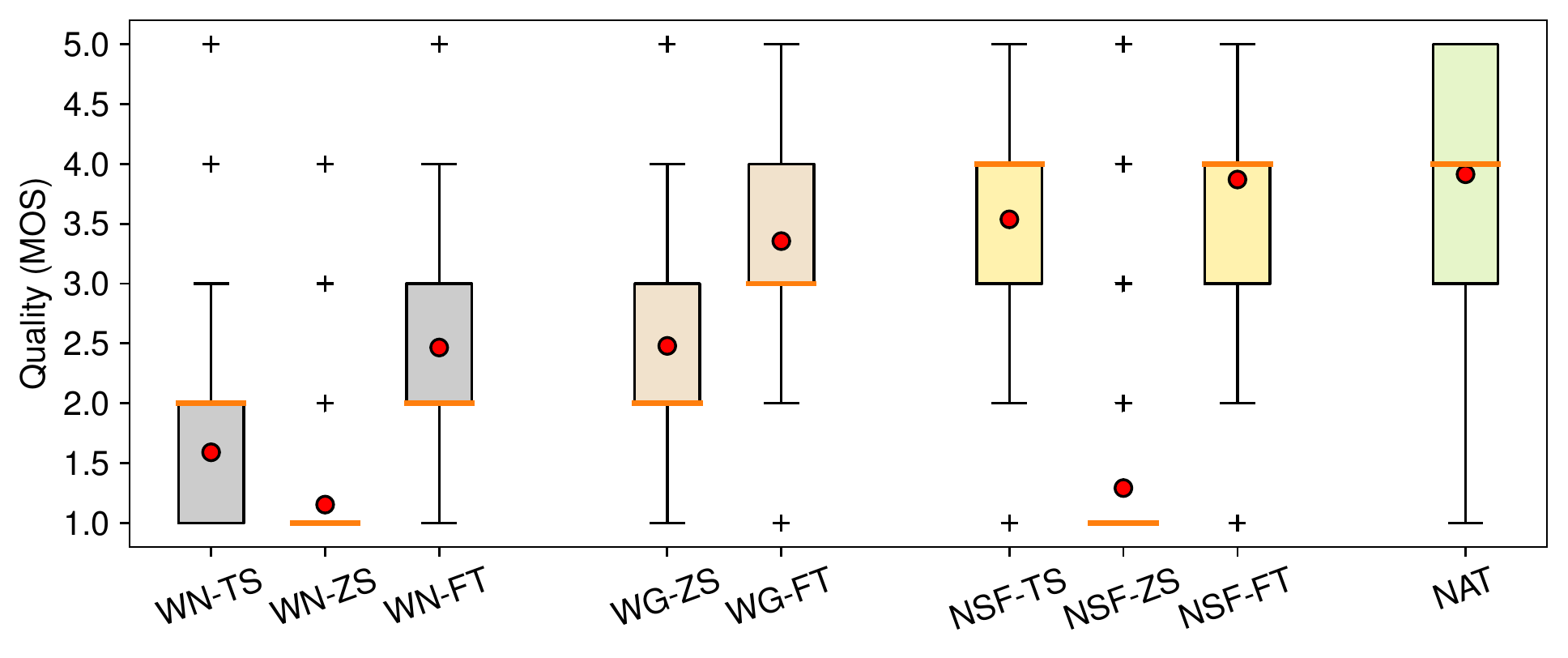}
\vspace{-5mm}
\caption{Box plots on naturalness evaluation results. Red dots represent mean score.} \label{fig:qua}
\end{figure}

\begin{figure}[thb]
 \centering
\includegraphics[width = 8.6cm]{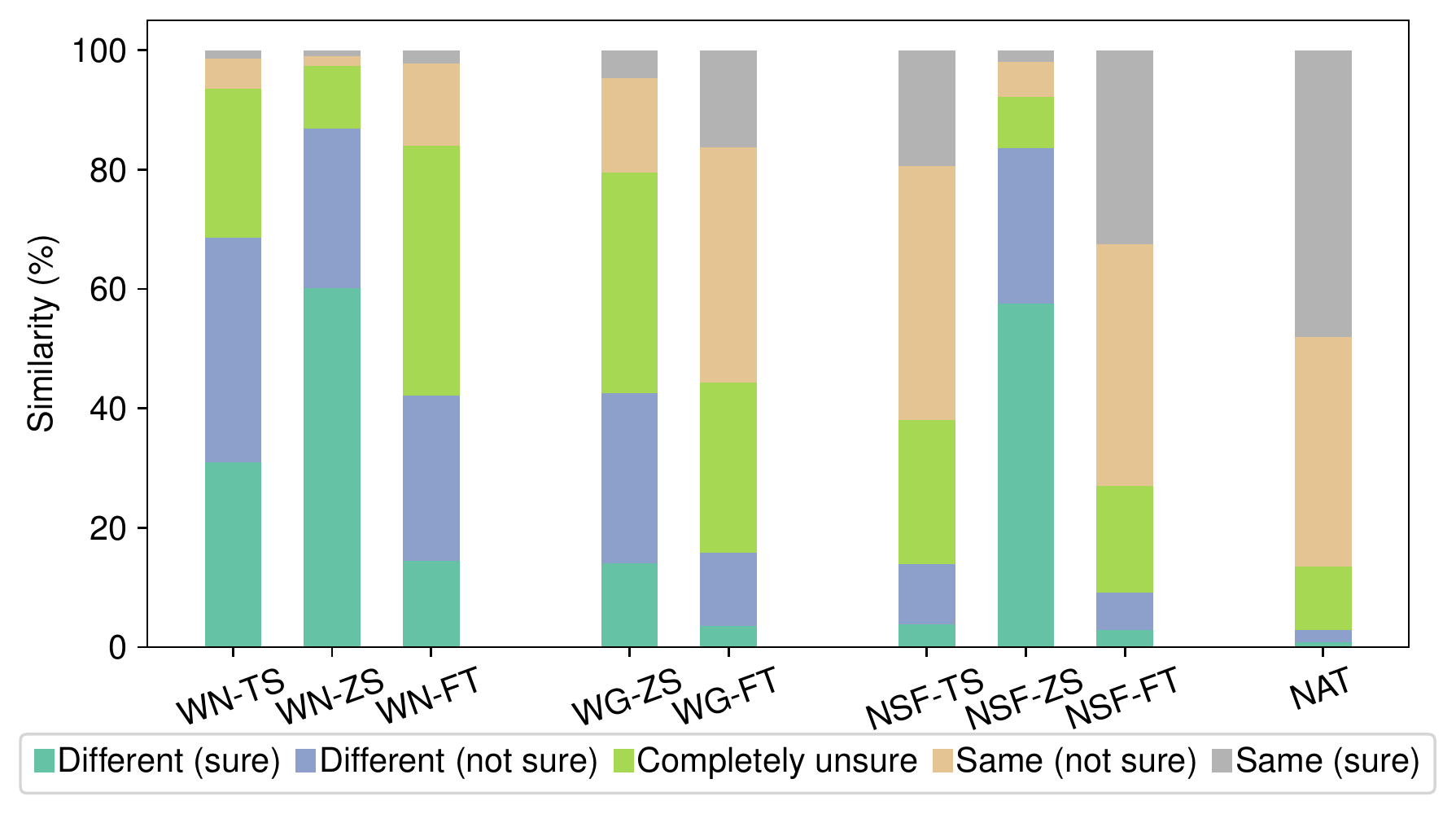}
\vspace{-5mm}
\caption{Evaluation results for similarity.} \label{fig:sim}
\end{figure}

Figures~\ref{fig:qua} and \ref{fig:sim} show the results of perceptual quality and similarity, respectively. T-tests with Holm-Bonferroni correction ($\alpha = 0.05$) were conducted for both evaluation metrics. In terms of quality, the pairwise difference was statistically significant except for the cases of \texttt{NSF-FT} vs. \texttt{NAT} ($p=0.18$) and \texttt{WN-FT} vs. \texttt{WG-ZS} ($p=0.78$). In terms of similarity, the pairwise difference was statistically significant except for \texttt{WN-FT} vs. \texttt{WG-ZS} ($p = 0.30$) and \texttt{WN-ZS} vs. \texttt{NSF-ZS} ($p=0.09$).  

These results demonstrate the effectiveness of the fine-tuning approach in the three types of waveform synthesizers. For WaveNet, Figure~\ref{fig:spec_rainbow} shows that the rainbow-gram\footnote{A 2-D figure showing the instantaneous frequency and spectral energy of every frequency bin in every analyzed frame \cite{engel2017neural}.} of the sample from \texttt{WN-FT} is less noisy than that from \texttt{WN-TS}. One possible reason is that \texttt{WN-FT} benefited from the pre-training on speech data and thus estimated waveform distributions with more confidence (i.e., smaller variance). \texttt{NSF-FT} generated music waveforms with less low-frequency energy than those from \texttt{NSF-TS}. In the case of WaveGlow, the training failed without fine-tuning, as mentioned previously. These results suggest that knowledge from the speech domain can be transferred to music signal modeling.

In the zero-shot scenario, Figure~\ref{fig:spec_rainbow} shows that \texttt{NSF-ZS} generated a noisy sample for trumpet and samples with an over-smoothed harmonic structure for violin. \texttt{WN-ZS} also produced samples with strong artefacts. One possible reason could be the mismatch of dynamic range of the input features in speech and music. Surprisingly, \texttt{WG-ZS} produced music waveforms with acceptable quality. The reason for this difference may stem from the property of inverse-AR flow, which deserves further analysis.

The comparison across the three types of neural waveform synthesizers demonstrates that \texttt{NSF-FT} performed best in both quality and similarity. Specifically, \texttt{NSF-FT}'s quality score was not significantly different from that of natural samples. The good performance of \texttt{NSF-FT} is also indicated by Figure~\ref{fig:spec_rainbow}: the rainbow-gram of \texttt{NSF-FT} is very close to those of natural speech, especially in the high-frequency band above note C7 for violin. Due to the assumption of the source-filter model, NSF may be more suitable to model the music signals of the string, woodwind, and brass instruments, where the sound has a regular harmonic structure. Such a model assumption may also help the NSF to learn from a small database, as the performance of \texttt{NSF-TS} suggests. In contrast, WaveNet and WaveGlow do not include the source-filter model assumption, and as such, they will need a sufficient amount of data to learn the periodicity in the music sounds. 


\section{Conclusions}
\label{sec:con}
In this paper, we compared the performance of WaveNet, WaveGlow, and NSF on the synthesis of musical instrument sounds in three scenarios: training from scratch on music data, zero-shot learning from speech data, and fine-tuning on pre-trained speech models. The results suggest that pre-training even on speech data could help to improve the performance in terms of music sound generation. 
Among the three synthesizers we examined, NSF performed best in both the fine-tuning and training-from-scratch scenarios.
Meanwhile, WaveGlow showed great potential in zero-shot cross-domain adaptation. 

For future work, we plan to evaluate the neural waveform synthesizers on other corpora such as NSynth~\cite{engel2017neural} and MAESTRO Dataset~\cite{hawthorne2018enabling}. By doing so, we can test the models on not only harmonic but also percussive sounds. We will also measure the impact of music data size on the training of neural waveform synthesizers.

\section{Acknowledgments}
This work was partially supported by a JST CREST Grant (JPMJCR18A6, VoicePersonae project), Japan, MEXT KANKENHI Grants (16H06302, 17H04687, 18H04120, 18H04112, and 18KT0051, 19K24371, 19K24373), Japan, and a Google AI Focused Research Awards Program, Japan. The numerical calculations were carried out on the TSUBAME 3.0 supercomputer at the Tokyo Institute of Technology.

The authors would like to thank Dr.\ Wei Ping from Baidu for his valuable comments. 

\vfill\pagebreak
\bibliographystyle{IEEEbib}
\bibliography{strings,refs}

\clearpage
\section{Appendix}
\begin{table}[htbp]
\caption{Pearson correlation coefficients (PCCs) and voiced/unvoiced (V/UV) decision errors  between manually corrected F0 and F0 extracted from generated sounds for each system.}
\begin{center}
\scriptsize
\setlength{\tabcolsep}{1pt}
\begin{tabular}{c|ccccccccc}
\hline\hline
&\texttt{WN-TS} & \texttt{WN-ZS} & \texttt{WN-FT} & \texttt{WG-ZS} & \texttt{WG-FT} &\texttt{NSF-TS} & \texttt{NFS-ZS} & \texttt{NSF-FT} & \texttt{NAT}\\
\hline
PCC&0.55 & –0.12 & 0.86 & 0.83 & 0.87 & 0.86 & –0.21 & 0.89 & 0.90 \\
\hline
V/UV (\%)&24.21 & 36.98 & 15.40 & 11.08 & 10.27 & 11.22 & 58.01 & 8.90 & 9.74 \\
\hline\hline
\end{tabular}
\end{center}
\label{tab:F0_corr_sys}
\end{table}
We extracted F0 of each utterance  using SPTK toolkit, and calculated the Pearson correlation coefficients (PCCs) between the extracted F0   and manually corrected  F0 for each system. We also calculated voiced/unvoiced (V/UV)   decision    errors     of each system based on the extracted F0.
Table~\ref{tab:F0_corr_sys} shows that there is an obvious mismatch between the manually corrected  F0 and automatically extracted F0 since PCCs of \texttt{NAT} is less than 1.
From Table~\ref{tab:F0_corr_sys} we can see that F0 extracted from system  \texttt{NSF-FT} has the highest correlation coefficient among all experimental systems, while \texttt{WN-ZS} and \texttt{NFS-ZS} get the lowest coefficients. Similar trends are detected in V/UV decision errors, and they are in consistent with the listening test results. 


\begin{table}[htbp]
\caption{Pearson correlation coefficients (PCCs) between manually corrected F0 and F0 extracted from generated sounds for each instrument in each system.}
\begin{center}
\scriptsize
\setlength{\tabcolsep}{1pt}
\begin{tabular}{c|ccccccccc}
\hline\hline
&\texttt{WN-TS} & \texttt{WN-ZS} & \texttt{WN-FT} & \texttt{WG-ZS} & \texttt{WG-FT} &\texttt{NSF-TS} & \texttt{NFS-ZS} & \texttt{NSF-FT} & \texttt{NAT}\\
\hline
Violin& 0.67 & –0.28 & \cellcolor {green}0.83 & 0.76 & 0.81 & 0.84 & –0.29 & 0.86 & 0.83 \\
Cello& 0.49 & 0.41 & 0.62 & 0.55 & 0.53 & 0.76 & 0.54 & 0.92 & 0.54 \\
Viola& 0.54 & 0.44 & 0.56 & 0.64 & 0.66 & 0.83 & \cellcolor {green}0.80 & 0.63 & 0.63 \\
Flute& \cellcolor {yellow}\textbf{–0.12} & \cellcolor {yellow}\textbf{–0.55} & 0.67 & 0.50 & 0.56 & 0.52 &\cellcolor {yellow}\textbf{ –0.35} & 0.61 & 0.77 \\
Oboe& 0.66 & –0.52 & 0.96 & 0.70 & \cellcolor {green}0.88 & \cellcolor {green}0.89 & 0.05 & \cellcolor {green}0.93 & \cellcolor {green}0.89 \\
Clarinet& 0.66 & 0.21 & 0.78 & 0.77 & 0.77 & 0.78 & –0.10 & 0.86 & 0.85 \\
Saxophone& 0.62 & 0.25 & 0.51 & 0.55 & 0.54 & 0.74 & 0.14 & 0.85 & 0.59 \\
Trumpet& 0.48 & –0.01 & 0.76 & 0.67 & 0.69 & 0.73 & –0.20 & 0.76 & 0.68 \\
Horn& 0.31 & 0.11 & 0.56 & \cellcolor {yellow}\textbf{0.20} &\cellcolor {yellow} \textbf{0.29} & \cellcolor {yellow}\textbf{0.46} & 0.31 & \cellcolor {yellow}\textbf{0.49} & \cellcolor {yellow}\textbf{0.42} \\
Trombone& \cellcolor {green}0.77 & \cellcolor {green}0.75 & \cellcolor {yellow}\textbf{–0.05} & \cellcolor {green}0.81 & 0.75 & 0.73 & 0.01 & 0.89 & 0.73 \\
Tuba& –0.01 & 0.64 & 0.30 & 0.64 & 0.48 & 0.53 & 0.36 & 0.74 & 0.73 \\
\hline
Overall&0.55 & –0.12 & 0.86 & 0.83 & 0.87 & 0.86 & –0.21 & 0.89 & 0.90 \\
\hline\hline
\end{tabular}
\end{center}
\label{tab:F0_corr_instru}
\end{table}

 Table~\ref{tab:F0_corr_instru} has shown the PCCs of F0 for each instrument of each system. 
From Table~\ref{tab:F0_corr_instru}, we can see that most of the lowest PCCs values occur with the instrument named flute and horn. 
On one hand,  three systems which show the lowest PCCs with flute also show poor performance in listening test. They were either trained from scratch or generated directly from speech models. However, after fine-tuning, PCCs valus can become much larger. 
According to table~\ref{tab:corpus} flute, flute owns the highest maximum F0 values as well as the largest F0 ranges ($\text{F0}_\text{max}$ - $\text{F0}_\text{min}$) among all instruments. 

\begin{table}[t]
\caption{Mean value of quality scores for each instrument in each system. The highest quality score for each system are marked with green background, while the lowest scores are in yellow background.}
\begin{center}
\scriptsize
\setlength{\tabcolsep}{1pt}
\begin{tabular}{c|ccccccccc}
\hline\hline
&\texttt{WN-TS} & \texttt{WN-ZS} & \texttt{WN-FT} & \texttt{WG-ZS} & \texttt{WG-FT} &\texttt{NSF-TS} & \texttt{NFS-ZS} & \texttt{NSF-FT} & \texttt{NAT}\\
\hline
Violin& 1.78 & 1.19 & 2.50 & 2.26 & 3.50 & 3.58 & 1.12 & 3.69 & 4.09 \\
Cello& 1.13 & 1.30 & 2.11 & 2.72 & 3.51 & 3.11 & 1.43 & 3.42 & 4.26 \\
Viola& 1.33 & \cellcolor{green}1.81 & 2.05 & 2.62 & 3.29 & 3.40 & \cellcolor{green}2.90 & 3.83 & 3.62 \\
Flute& 1.57 & \cellcolor {yellow}\textbf{1.00} & 2.74 & 2.45 & 3.65 & 3.84 & 1.09 & 4.00 & 4.03 \\
Oboe& 1.38 & \cellcolor {yellow}\textbf{1.00} & 2.51 & 1.74 & 3.25 & 3.47 & 1.11 & 3.92 & 3.89 \\
Clarinet& 1.64 & 1.34 & 2.57 & 2.86 & \cellcolor{green}3.76 & \cellcolor{green}3.88 & 1.07 & 4.09 & \cellcolor{green}4.29 \\
Saxophone& 1.73 & 1.13 & 2.47 & \cellcolor{green}2.87 & 3.13 & 3.40 & 1.10 & 3.75 & 3.70 \\
Trumpet& \cellcolor{green}1.94 & 1.04 & \cellcolor{green}2.84 & 2.49 & 3.60 & 3.69 & 1.11 & \cellcolor{green}4.17 & 4.07 \\
Horn& 1.70 & 1.04 & 2.74 & \cellcolor{green}2.87 & 3.37 & 3.57 & 1.09 & 3.87 & 3.61 \\
Trombone& 1.07 & 1.31 & 1.93 & 2.69 & 3.02 & 3.37 & \textbf{1.04} & 3.67 & 3.54 \\
Tuba& \cellcolor {yellow}\textbf{1.06} & 1.19 & \cellcolor {yellow}\textbf{1.13} & \cellcolor {yellow}{\textbf{1.72}} & \cellcolor {yellow}\textbf{1.69} & \cellcolor {yellow}\textbf{2.70} & 2.59 & \cellcolor {yellow}\textbf{3.41} &
\cellcolor {yellow}\textbf{3.20} \\
\hline
Overall & 1.48 & 1.21 & 2.33 & 2.48 & 3.25 & 3.46 & 1.42 & 3.80 & 3.85 \\
\hline\hline
\end{tabular}
\end{center}
\label{tab:qua_instru}
\end{table}

\begin{table}[htbp]
\caption{Mean value of similarity scores for each instrument in each system. The highest similarity score for each system are marked with green background, while the lowest scores are in yellow background}
\begin{center}
\scriptsize
\setlength{\tabcolsep}{1pt}
\begin{tabular}{c|ccccccccc}
\hline\hline
&\texttt{WN-TS} & \texttt{WN-ZS} & \texttt{WN-FT} & \texttt{WG-ZS} & \texttt{WG-FT} &\texttt{NSF-TS} & \texttt{NFS-ZS} & \texttt{NSF-FT} & \texttt{NAT}\\
\hline
Violin& 2.32 & 1.52 & 2.81 & 2.16 & 3.38 & 3.82 & 1.55 & 4.03 & 4.34 \\
Cello& 1.41 & 1.91 & 2.24 & 2.52 & 3.44 & \cellcolor {yellow}2.74 & 1.87 & \cellcolor {yellow}2.63 & \cellcolor {yellow}3.94 \\
Viola& 2.17 & \cellcolor{green}2.31 & 2.55 & 3.26 & 3.67 & 3.38 & \cellcolor{green}3.29 & 3.21 & 4.07 \\
Flute& 2.09 & 1.21 & 2.80 & 2.40 & 3.60 & 3.62 & 1.38 & 3.89 & 4.33 \\
Oboe& 1.73 & \cellcolor {yellow}1.04 & 2.33 & \cellcolor {yellow}1.94 & 2.96 & 3.62 & \cellcolor {yellow}1.31 & 4.08 & 4.25 \\
Clarinet& 2.00 & 1.55 & 2.43 & 2.68 & 3.53 & 3.67 & 1.42 & 4.03 & 4.18 \\
Saxophone& 2.33 & 1.70 & 2.83 & \cellcolor{green}3.50 & 3.87 & 3.52 & 1.40 & 4.14 & 4.31 \\
Trumpet& 2.39 & 1.41 & 2.80 & 2.58 & 3.63 & 3.85 & 1.55 & \cellcolor{green}4.25 & 4.42 \\
Horn& \cellcolor{green}2.64 & 1.75 & \cellcolor{green}2.89 & 3.40 & \cellcolor{green}4.02 & \cellcolor{green}4.00 & 1.51 & 4.21 & 4.40 \\
Trombone& 1.38 & 1.89 & 2.40 & 3.02 & 3.45 & 3.85 & 1.64 & 3.92 & 4.25 \\
Tuba& \cellcolor {yellow}1.28 & 2.08 & \cellcolor {yellow}1.62 & 3.00 & \cellcolor {yellow}2.85 & 3.23 & 3.17 & 3.81 & \cellcolor{green}4.51 \\
\hline
Overall & 1.98 & 1.67 & 2.52 & 2.77 & 3.49 & 3.57 & 1.83 & 3.84 & 4.27 \\
\hline\hline
\end{tabular}
\end{center}
\label{tab:sim_instu}
\end{table}

Table~\ref{tab:qua_instru} and~\ref{tab:sim_instu} show the quality and similarity scores for each instrument of each system, respectively. 
In table~\ref{tab:qua_instru}, tuba is blamed for the lowest MOS scores in most experimental systems as well as natural sound.

\end{document}